\documentclass[letterpaper]{article} 
\usepackage{aaai2026}
\usepackage{times}  
\usepackage{helvet}  
\usepackage{courier}  
\usepackage[hyphens]{url}  
\usepackage{graphicx} 
\usepackage{natbib}  
\usepackage{caption} 
\usepackage{amsmath}   
\usepackage{booktabs}  

\usepackage{algorithm}
\usepackage{algorithmic}
\usepackage[most]{tcolorbox}

\usepackage{newfloat}
\usepackage{listings}
\DeclareCaptionStyle{ruled}{labelfont=normalfont,labelsep=colon,strut=off} 
\floatstyle{ruled}
\newfloat{listing}{tb}{lst}{}
\floatname{listing}{Listing}
\title{MCTS-SQL: Light-Weight LLMs can Master the Text-to-SQL through Monte Carlo Tree Search}
\author{
   Shuozhi Yuan\textsuperscript{\rm 1}, Limin Chen\textsuperscript{\rm 1}, Miaomiao Yuan\textsuperscript{\rm 2}, Jin Zhao\textsuperscript{\rm 1}
}
\affiliations{
    \textsuperscript{\rm 1}China Telecom Digital Intelligence\\
     \textsuperscript{\rm 2}Institute of Computing Technology, Chinese Academy of Sciences\\

    Beijing, China\\
    yuansz@chinatelecom.cn
}

\begin{document}

\maketitle

\begin{abstract}
Text-to-SQL is a fundamental yet challenging task in the NLP area, aiming at translating natural language questions into SQL queries. While recent advances in large language models have greatly improved performance, most existing approaches depend on models with tens of billions of parameters or costly APIs, limiting their applicability in resource-constrained environments. For real world, especially on edge devices, it is crucial for Text-to-SQL to ensure cost-effectiveness. Therefore, enabling the light-weight models for Text-to-SQL is of great practical significance. However, smaller LLMs often struggle with complicated user instruction, redundant schema linking or syntax correctness. To address these challenges, we propose \textbf{MCTS-SQL}, a novel framework that uses Monte Carlo Tree Search to guide SQL generation through multi-step refinement. Since the light-weight models' weak performance of single-shot prediction, we generate better results through several trials with feedback.  However, directly applying MCTS-based methods inevitably leads to significant time and computational overhead. Driven by this issue, we propose a token-level \textbf{prefix-cache mechanism} that stores prior information during iterations, effectively improved the execution speed. Experiments results on the SPIDER and BIRD benchmarks demonstrate the effectiveness of our approach. Using a small open-source Qwen2.5-Coder-1.5B, our method outperforms ChatGPT-3.5. When leveraging a more powerful model Gemini 2.5 to explore the performance upper bound, we achieved results competitive with the SOTA. Our findings demonstrate that even small models can be effectively deployed in practical Text-to-SQL systems with the right strategy.
\end{abstract}


\begin{figure}[t]
  \begin{minipage}{\linewidth}
    \centering
    \includegraphics[width=\linewidth]{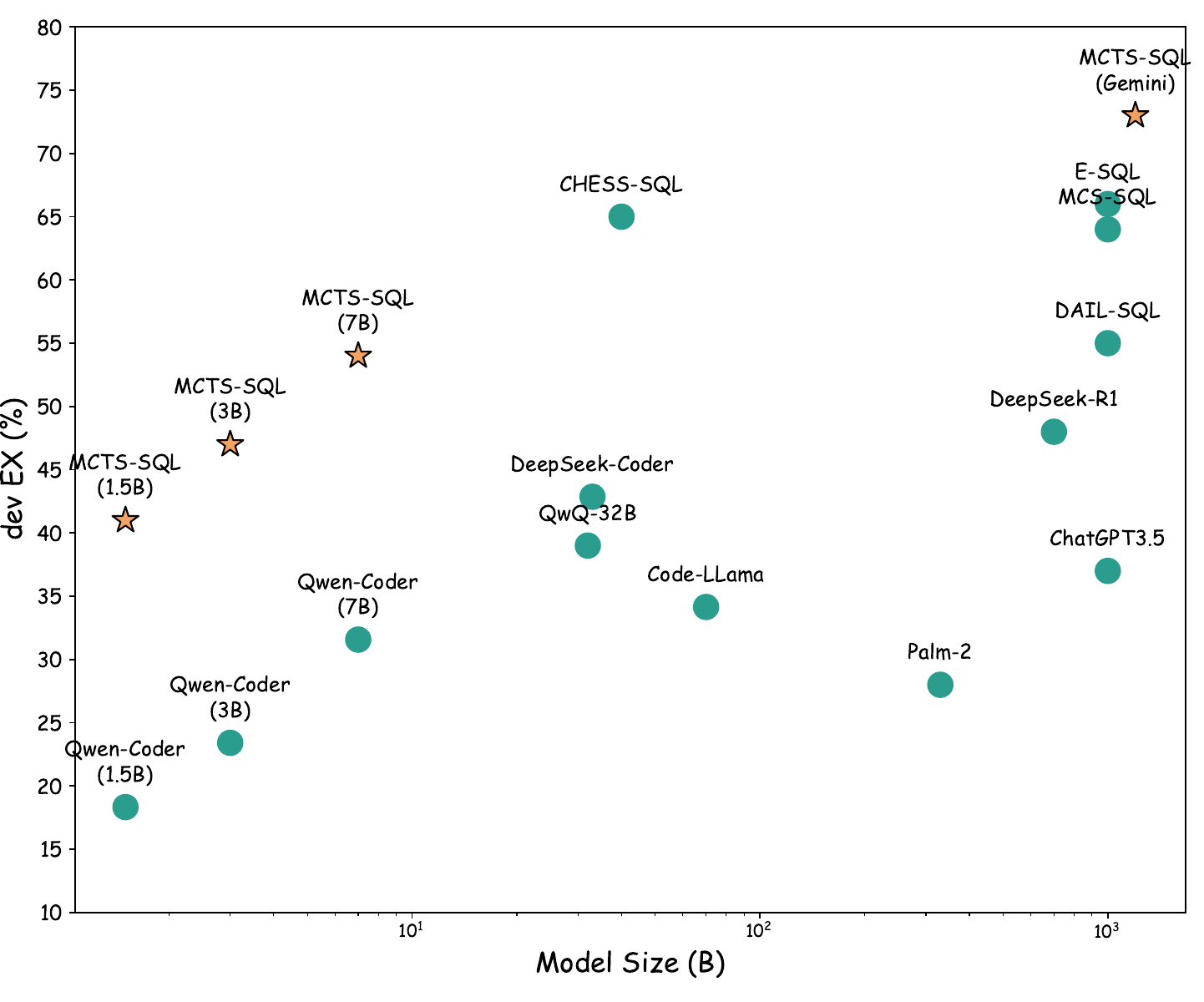}
    \caption{Execution accuracy comparison of MCTS-SQL across some existing methods. The results show that MCTS-SQL significantly enhances the performance of light-weight models, achieving performance comparable to some larger models. And when using Gemini2.5 as the base model, we achieve results competitive with the SOTA.}
    \label{Fig1}
  \end{minipage}
\end{figure}

\section{Introduction}

Text-to-SQL is a task aimed at converting natural queries into SQL, which plays a critical role in data analytics and supports a wide range of real-world applications \cite{Survey,2024survey}. Recent advances in LLM \cite{Gemini, GPT4} have significantly improved the performance of Text-to-SQL systems. However, most of these powerful methods \cite{MAC-SQL,MCS,ESQL} rely on extremely huge models or costly APIs, making them expensive and can not be used in resource-constrained environments. For real-world use, especially on edge devices, running efficiently is crucial to keep low resources cost. This raises an important challenge: \textbf{How can we enable lightweight models to effectively handle Text-to-SQL tasks ?}

Most common errors made by these small LLMs is the mistake understanding of users' intent, wrong schema selection and syntax errors. Due to the poor performance of single-shot prediction, an intuitive way to address these challenges is to conduct a trial-and-feedback mechanism to iteratively optimize the generated SQL. However, feedback without any constraints or guidance is inefficient. Therefore, a more powerful optimization strategy is needed to direct better solutions. Monte Carlo Tree Search (MCTS) has proven to be an efficient tool in decision-making and optimization tasks. Recent studies \cite{Pitanov, li2023, chen2024} have demonstrated that MCTS can be effectively applied to problems requiring iterative improvements. Given its strengths, MCTS presents a practical tool for optimizing SQL generation in Text-to-SQL tasks.

In this paper, we introduce MCTS-SQL, a practical approach that effectively applies MCTS to the Text-to-SQL task. Every components of our design follows one principle: \textbf{progressively reducing the search space to align with the limited reasoning capacity of light models.} 

To realize this, we introduce three components:(i)\textbf{Selector} prunes irrelevant schema elements, reducing prompt complexity;
(ii)\textbf{Direct Generator} provides a strong initial SQL candidate, avoiding deep search from scratch;(iii)\textbf{MCTS-Refiner} iteratively improves the SQL through guided exploration.

While effective, MCTS-based approaches introduce a new challenge: computation cost. Compared to single-shot prediction, iterative refinement inevitably increases token usage and inference time. Rethinking the whole pipeline of MCTS-SQL, we find that a large amount of inputs of every instructions are highly repetitive (e.g., database schema, field descriptions, few-shot examples). To address the latency from re-computing these inputs, we design a \textbf{prefix-cache mechanism}. Repeated inputs are computed once and cached, during MCTS iterations, only changing feedback and refinements are processed. This approach reduces inference time by \textbf{53\%}.

We evaluate the performance of MCTS-SQL on two widely-used benchmarks: The Spider \cite{SPIDER} and BIRD \cite{BIRD}. The results show that MCTS-SQL based on the Qwen-2.5-Coder-Instruct-1.5B \cite{Qwen} outperforms ChatGPT-3.5, and 3B version achieves even better performance than some earlier GPT-4o based methods. Moreover, to explore the boundaries of our algorithm, we evaluate results based on Gemini-2.5. Unsurprisingly, we achieve competitive results, with execution accuracy of \text{72.91\%} on BIRD. The comparison across some existing methods can be seen in Figure 1.

The main contributions of our proposed MCTS-SQL can be summarized as follows:

\begin{itemize}

\item We propose \textbf{MCTS-SQL}, which apply Monte Carlo Tree Search to Text-to-SQL, demonstrating that small models can effectively handle this complex task.

\item We design a search-space-reducing pipeline with Selector, Direct Generator, and MCTS-Refiner to guide SQL optimization.

\item We introduce a novel \textbf{prefix-cache} mechanism to reduce redundant computation, significantly improving inference efficiency.

\item Experiments on BIRD and SPIDER show that MCTS-SQL enables small models to achieve practical accuracy. For instance, using Qwen-1.5B, our method reaches \textbf{40.69\%} EX on BIRD dev set. When leveraging a stronger model Gemini 2.5, it achieves \textbf{72.91\%} EX on BIRD dev, which is competitive with the current SOTA.

\end{itemize}

\begin{figure*}[htbp]
  \begin{center}
  \includegraphics[width= 6.7in]{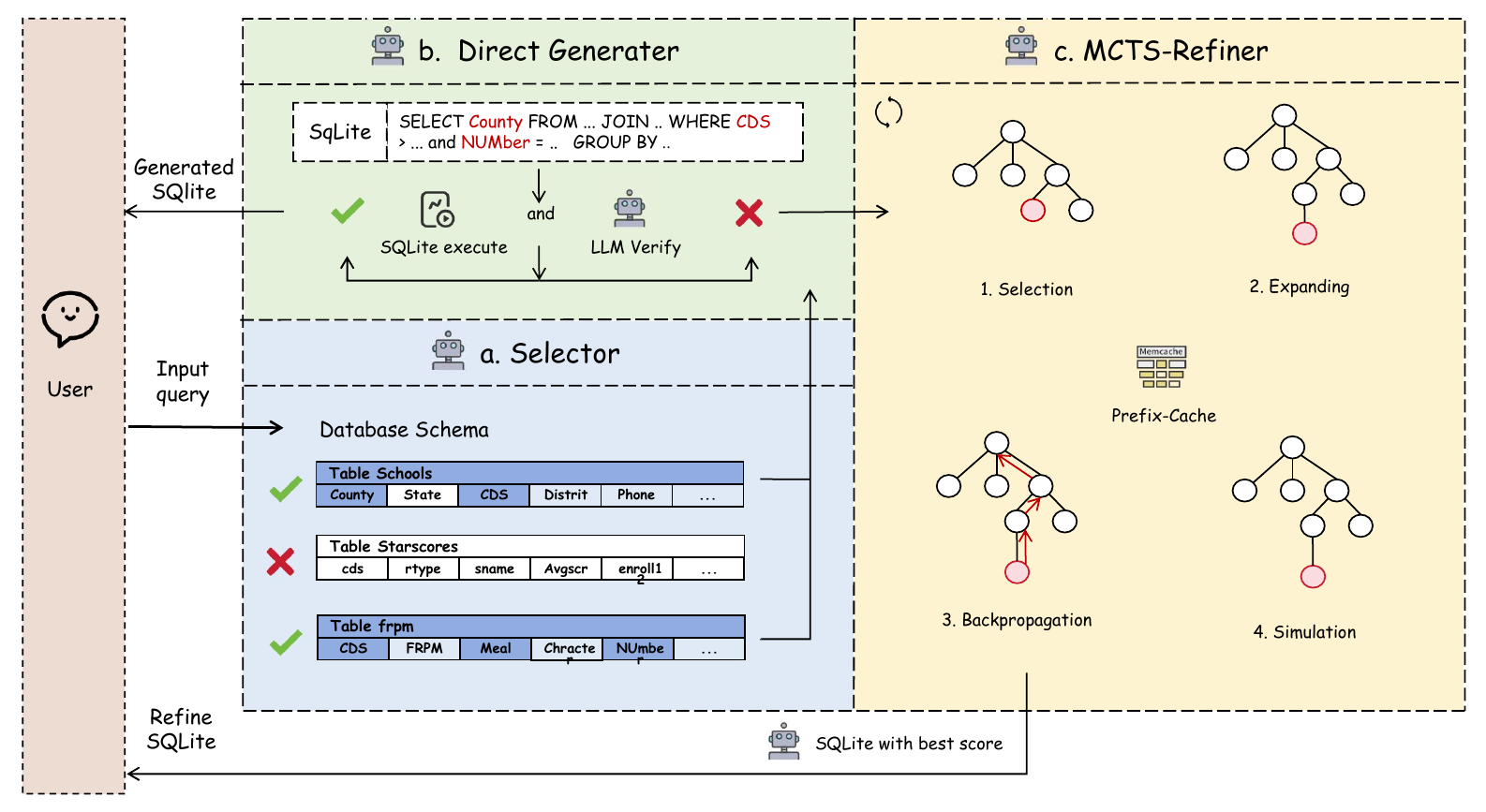}
  \caption{The MCTS-SQL framework consists of three core components: the \textbf{Selector}, the \textbf{Direct Generator} and the \textbf{MCTS-Refiner}. The Selector is used to filter the most relevant tables and columns based on the user's intent. The Direct Generator aims to produce an initial SQL query. And the MCTS-Refiner is activated when the initial SQL query fails both execution and LLM-based verification checks. which adopts iterative trial-and-feedback optimization to refine the query progressively. }
  \label{Fig2}
  \end{center}
\end{figure*}

\section{Related Work}
In this section, we provide an overview of related work on Text-to-SQL and Monte Carlo Tree Search, highlighting their relevance and main differences to our proposed research.

\subsection{Text-to-SQL}
Text-to-SQL aims to bridge natural language queries and structured database queries, and numerous approaches are proposed to address its challenges. Early systems, such as LUNAR \cite{LUNAR} and NaLIX \cite{NaLIX}, employed rule-based methods that manually crafted grammar rules and heuristics. However, the generalization performance of these methods across different tasks or databases is difficult to guarantee.

The deep learning marked a turning point for Text-to-SQL. End-to-end models like Seq2SQL \cite{Seq2SQL} and SQLNet \cite{SQLNet} directly mapped natural language to SQL but struggled with complex queries, especially those involving nested structures or intricate reasoning. Pre-trained Language Models (PLMs), such as TaBERT \cite{TaBERT} and BERT-SQL \cite{BERT-SQL}, enhance cross-domain generalization and improve the accuracy of SQL generation. However, these methods require a certain amount of domain-specific SQL training data, which makes them difficult to land in practical applications 

Recently, Large Language Models (LLMs) such as GPT-4 \cite{GPT4}, Palm-2 \cite{Palm}, and LLaMA \cite{llama} have revolutionized Text-to-SQL tasks. These methods excel in zero-shot and few-shot settings without any extensive training data.\cite{MCS,Chess,ESQL}. DAIL-SQL \cite{DAIL-SQL} optimized prompt engineering, focusing on question representation, prompt structure, and example selection to enhance SQL accuracy with minimal supervision. Additionally, MAC-SQL \cite{MAC-SQL} introduced a collaborative framework integrating decomposer, auxiliary selector, and refiner modules for iterative SQL refinement. Recently, Xiyan-SQL\cite{Xiyan}, Chase-SQ\cite{chase}L, and DSAIR-SQL\cite{DSAIR} have also achieved improvements in NL2SQL performance through fine-tuning and sophisticated agent engineering. The most similar works, Alpha-SQL\cite{Alpha} and SQL-o1\cite{SQL-o1}, perform heuristic, action-level search from scratch, whereas our MCTS-SQL adopts a progressive refinement strategy and leverages a prefix-cache mechanism to minimize computation during iterative exploration.

From the analysis of existing methods, it becomes clear that advancing Text-to-SQL performance relies on the models' understanding and reasoning capabilities. However, these models are often of large scale and costly, making them impractical for real-world resource-constrained application. 
Our core motivation is to enable lightweight models to achieve practical performance in Text-to-SQL tasks. To this end, we incorporate Monte Carlo Tree Search (MCTS) to guide the generation.

\subsection{Monte Carlo Tree Search}

MCTS is widely used for planning complex problems, and a large number of downstream experiments have demonstrated its effectiveness.
For example, \cite{Pitanov} demonstrates its benefits in multi-agent path search, highlighting the advantages over traditional heuristic search methods. Similarly, \cite{li2023} use MCTS to effectively address various types of SAT problems. Recently,  combining MCTS with large-scale language models has been a great trend. \cite{chen2024} proposed IMCTS, an approach designed to enhance the mathematical reasoning capabilities of fine-tuned LLMs. \cite{xu2023} integrated MCTS with a lightweight energy function, demonstrating notable performance improvements. In addition, MCTSr \cite{MCTSr} introduced systematic exploration and heuristic self-refinement mechanisms, further advancing its applications in complex decision-making tasks.

Building on these successes, our work is the first to introduce Monte Carlo Tree Search into the Text-to-SQL domain.The core idea is very simple: reduce errors through iterative trial and error. However, naive exhaustive attempts are inefficient and impractical.  To address this, we leverage MCTS to find a more efficiently and reliable exploration path.  Moreover, we design a prefix-cache to avoid repeated computational  overlead.

\section{MCTS-SQL Framework}

As shown in Figure 2, the MCTS-SQL framework consists of three key components: the Selector, Direct Generator, and MCTS-Refiner. The Selector filter relevant tables and schema elements based on the user query, while the Direct Generator produces an initial SQL query. Queries that fail or yield errors are refined by the MCTS-Refiner through iterative tree search. A detailed explanation of each component is provided in the subsequent section. The collaboration process of our MCTS-SQL is presented in Algorithm 1.

\subsection{Schema}
Before introducing the specific components, we would like to describe the special design of effectively translating database structures. Combining the database schema information in the prompt is essential for enabling the LLM to comprehend the database structure accurately and generate precise queries. We present a novel method that illustrates the hierarchical relationships between databases, tables, and columns using a semi-structured format.  

To be specific, we provide the table name and corresponding description for each table(which can be omitted if not necessary). The table information is converted into a list where each entry is a tuple containing a column of details. Each column includes the name, data type, description, and example values, thus providing a comprehensive view of its contents. In addition, foreign keys must be included to represent the relationships between tables accurately. Understanding hierarchical relationships is critical for query generation.  All the agents in this paper introduce database information through this schema.

\begin{algorithm}
    \caption{The algorithm of MCTS-SQL}
    \textbf{Input}: query q, database schema db, knowledge kg\\
    \textbf{Output}: SQL statement
    \begin{algorithmic}[1] 
        \STATE db' = {$LLM_{Selector}$}(q, db, kg)
        \STATE sql,err = {$LLM_{Direct Generator}$}(q, db, kg)
        \STATE ver = {$LLM_{Verifier}$}(sql,q,db,kg,)
        \IF{err is NULL and ver is ok}
        \STATE \textbf{return} sql
        \ELSE
        \STATE count = 0
        \WHILE{count $<$ maxRollout}
        \STATE select a node
        \STATE cri = {$LLM_{Critiquer}$}(sql, err, q, db, kg)
        \STATE ref = {$LLM_{Refiner}$}(sql, err, q, db, kg, cri)
        \STATE score = {$LLM_{Evaluater}$}(ref, err, q, db, kg)
        \STATE back-propagation
        \STATE update the UCT value
        \ENDWHILE
        \STATE sql = ref with best score
        \STATE \textbf{return} sql
        \ENDIF
    \end{algorithmic}
\end{algorithm}


\subsection{Selector}

The role of the Selector can be formally described as follows. Given an input triplet $\mathcal{X} = (Q, S,\mathcal{K})$, where $Q$ is the query, $S = {T, C}$ is the database schema consisting of tables ($T$) and columns ($C$), and $\mathcal{K}$ denotes the knowledge provided. The Selector aims to identify a minimal subset of tables and columns, denoted as $S' = {T', C'}$, which are necessary to answer the query $Q$. The behavior of the Selector is formally defined as follows:

\begin{equation}
S' = f_{\text{Selector}}(Q, S, \mathcal{K} \mid \mathcal{M})
\label{Eq1}
\end{equation}

Where $f_{\text{Selector}}(\cdot \mid \mathcal{M})$ represents the Selector's function, implemented via prompt engineering powered by a large language model $\mathcal{M}$.

The Selector serves as a schema pruning module that leverages a large language model to identify tables and columns relevant to the query. Given $(Q, S, \mathcal{K})$, it interprets the query and associated knowledge to produce a focused subset $S' = {T', C'}$ necessary for answering $Q$. This process eliminates irrelevant schema elements, reducing the noise of schema and preventing prompt overflow caused by the full schema. By reducing the context, the Selector improves both the accuracy and efficiency of subsequent SQL generation.

\subsection{Direct Generator}
The purpose of the Direct Generator is to generate SQL queries directly through an end-to-end process. It can be described as follows, where $R$ represents the generated SQL query.

\begin{equation}
R = f_{\text{Direct Generator}}(Q, S', \mathcal{K} \mid \mathcal{M})
\label{Eq2}
\end{equation}

After the SQL is generated, it follows two steps of evaluation. First, an executor checks its syntactic correctness and successful execution. Then, an LLM verifies if the SQL meets the user's requirements. The LLM-based verifier can be formalized as:

\begin{equation}
V = f_{\text{Verifier}}(R, Q, S', K\mid\mathcal{M})
\end{equation}

Specifically, the Direct Generator employs chain-of-thought prompting. \cite{COT}  We assemble the relevant table and field information obtained from the Selector mentioned above with the user input. The LLM processes this input to generate SQL queries, accompanied by a detailed rationale. Additionally, we employ a few-shot learning strategy, using several in-context examples to improve the LLM's understanding of task-specific instructions and enhance its generalization capabilities.

\subsection{MCTS-Refiner}

Typically, SQL queries generated by the Direct Generator fail to meet task requirements due to syntactic errors or mismatch with the user's intent. The MCTS-Refiner aims to refine SQL queries using the self-critique mechanism to optimize the query iteratively. The MCTS-Refiner is conditionally activated via two checks: (1) Execution Failure: If the generated SQL fails to run. (2) Semantic Error: If execution succeeds but an LLM-based Verifier identifies user's intent  mismatch.The main workflow of the proposed method consists of several stages, detailed as follows:

\textbf{Initialization:} The root node is initialized with the suboptimal SQL generated by the Direct Generator as a reference for step-by-step optimization to reduce the complexity of the search process.

\textbf{Selection:} Following the existing practices, we define a function $P$ to rank all generated SQL queries that are not fully expanded. The node with the highest value is selected for further refinement. The function $P$ of a node $a$ can be defined as follows, where $r_a$ represents the set of results associated with node $a$. 

\begin{equation}
P(a) = \frac{1}{2} \left( \min r_a + \frac{1}{|r_a|} \sum_{i=1}^{|r_a|} r_a^i \right)
\end{equation}

\begin{figure}
  \begin{center}
  \includegraphics[width= 3.2in]{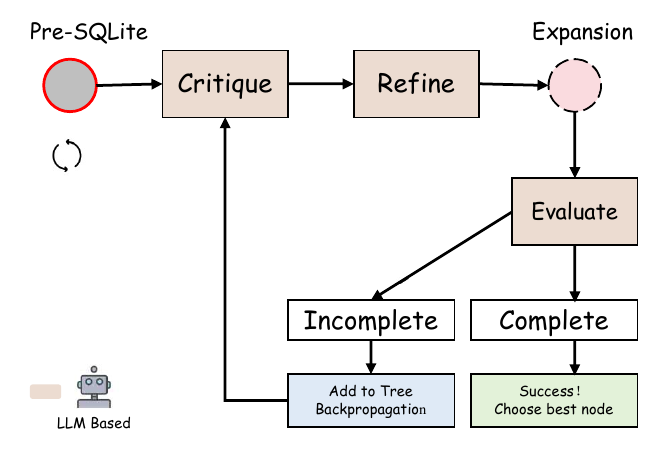}
  \caption{The main workflow of our proposed MCTS-refiner. The SQL generated in the last step is firstly get a critique. Then, based on the critique, a refinement is provided. The search tree is now expanded.If the iteration is complete, the node with best score is selected as the final output, otherwise, the node will be added to the search tree and backpropagation.}
  \label{Fig3}
  \end{center}
\end{figure}

\textbf{Self-Refine:}
The SQL query $a$ is initially executed by the executor to get the error information $E_{a}$, which is then used to refine the query through the self-refine framework. In this process, the LLM generates a critique $c$, serving as the guidance for refining the query and producing an improved SQL query $a'$. Specifically, $E_{a}$ represents the error details related to the initial SQL query $a$, and $I$ denotes the prompt used in the Direct Generator, which includes the input query $Q$, the database schema $S'$, and relevant knowledge $K$. The details can be formally described as follows:

\begin{equation}
c = f_{\text{Critiquer}}(a, E_{a}, I\mid\mathcal{M})
\end{equation}
\begin{equation}
a' = f_{\text{Refiner}}(a,c, E_{a}, I\mid\mathcal{M})
\end{equation}

The Self-refine module designed a refinement mechanism using error feedback and critique generation to enhance the accuracy and robustness of SQL queries.

\textbf{Self-Evaluation:}
The refined SQL query is evaluated to obtain a reward value, denoted as $r$, and its corresponding $P$-value is computed. To be specific, we proposed a model-based self-reward feedback mechanism, with the reward value constrained within the range of -95 to 95. To ensure the reliability and fairness, the highest scores are deliberately suppressed. The reward $r$ is formally defined as:

\begin{equation}
r_a = f_{\text{Evaluater}}(a',E_{a'}, I\mid\mathcal{M})    
\end{equation}

\textbf{Backpropagation:}
The value $r$ of the refined SQL query is back-propagated through the search tree, updating the value information of the parent node and other relevant nodes. If the $P$-value of any child node is changed, the corresponding $P$-value of its parent node is recalculated accordingly. The process can be described as follows:

\begin{equation}
P'(a) = \frac{1}{2} \left( P(a) + \max_{i \in \text a.children} P(i) \right)
\end{equation}

\textbf{UCT update:}
Following the existing practice \cite{MCTSr}, after updating the $P$ values for all nodes,  we choose the \textit{UCT} function to measure the combined value of each node, which is used as an important basis for expansion in the next selection stage. The \textit{UCT} value of a node $a$ is formally defined as:

\begin{equation}
UCT_a = P(a) + c \sqrt{\frac{\ln N(\text{Father}(a)) + 1}{N(a) + \epsilon}}
\end{equation}

In this formulation, $N(.)$ denotes the total number of visits to a given node, and $c$ is a constant that balances the trade-off between $P$-value and visit times. The term $\epsilon$ is a small constant to prevent division by zero.

The algorithm proceeds through all these steps iteratively until the maximum rollout numbers are reached. And the SQL queries with the highest score $r$ is chosen as the final output. 

\subsection{Prefix-Cache}

Rethinking the MCTS iterations, we observed that a large part of input prompts remain unchanged, such as database schema, field descriptions, and few-shot examples. Re-computing these repeated components introduces unnecessary latency and computational overhead.

To address this issue, we design a prefix-caching mechanism (as can be seen in Fig.4) that reuses the intermediate K/V states of the transformer decoder layers:

\begin{itemize}
\item First Pass: When the model processes an input prefix for the first time, we cache the K/V states generated for decoder layer.

\item  Subsequent Passes: If the same prefix appears again, the cached K/V states are restored to skip redundant computation and focus on the new suffix.

\end{itemize}

In practice, we define the prefix as the invariant components(e.g.,field descriptions, few-shot examples), while the suffix consists of the parts that change during iterations, such as instructions, evaluations and execution feedback. To align with the prefix-caching mechanism, the overall prompt is structured as: dataset schema + few-shot demonstrations + specific instructions + feedback.

\begin{figure}
  \begin{center}
  \includegraphics[width= 3.2in]{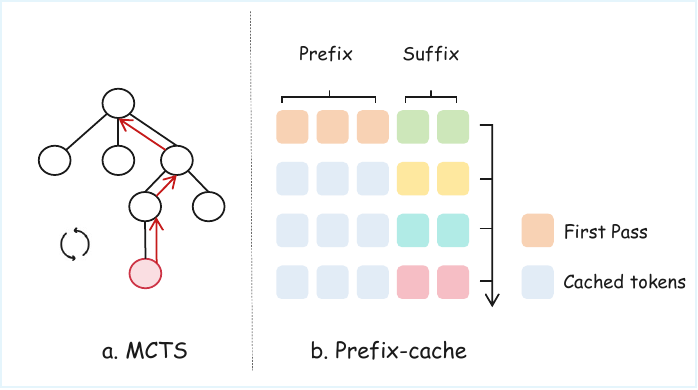}
  \caption{Illustration of the proposed optimization strategies. (a) MCTS-based iterative refinement for SQL generation. (b) Prefix-caching mechanism that reuses cached K/V states for invariant prompt components, reducing redundant computation and improving efficiency during multiple iterations.}
  \label{Fig4}
  \end{center}
\end{figure}

\section{Experiments}
To evaluate the performance of our MCTS-SQL, we present the implementation details, explain the experiments performed, and offer a thorough analysis of the results. We evaluate our MCTS-SQL frameWORK using two Text-to-SQL benchmarks: Spider and BIRD.


\begin{table}[htbp]
\centering
\begin{footnotesize} 
\begin{tabular}{lccc}
\toprule
\textbf{Method} & \textbf{dev EX} &\textbf{test EX} &\textbf{dev VES}\\
\midrule
Palm-2 & 27.38 & 33.04 & -\\
ChatGPT-3.5 & 36.64 & 40.08 & 42.30\\
DIN-SQL+GPT-4 & 50.72 & 55.09 & 58.79 \\
DAIL-SQL+GPT-4 & 54.76 & 57.41 & 56.08 \\
MAC-SQL+GPT-4 & 59.39 & 59.59 & 66.39 \\
MCS-SQL+GPT-4 &63.36 &65.45 &61.23\\
Sql-o1 &63.4& 64.8&-\\
CHESS &65.00 &66.69&62.77\\
ByteBrain &65.45&68.87&-\\
ASKData+GPT-4o &65.19&65.62&60.25\\
E-SQL+GPT-4o & 65.58&66.29&62.43\\
Alpha-SQL & 69.70& -& -\\
XiYan-SQL & 73.34&75.63&-\\
Chase-SQL &74.46&74.79&-\\
DSAIR-SQL &74.32&74.12&-\\

\midrule
\textbf{Ours+Qwen-1.5B}&\textbf{40.69}& \textbf{43.72}&\textbf{44.87}\\
Ours+Qwen-3B&46.71&48.37&48.19\\
Ours+Qwen-7B&53.61&51.79&52.21\\
\midrule
Ours+GPT-4o-mini&63.15&61.39&60.78\\
Ours+GPT-4o&69.40&68.91&66.24\\
Ours+Gemini2.5-pro&72.91&74.74&72.66\\
\bottomrule
\end{tabular}
\end{footnotesize}
\caption{Comparison with the results of existing methods on BIRD of the Execution accuracy and Valid Efficiency Score.The Qwen models in this table are Qwen-Coder-Instruct.}
\end{table}

\subsection{Evaluation Metrics}
To evaluate our proposed method's performance, we use two metrics: Execution Accuracy(EX) and Valid Efficiency Score(VES).\cite{BIRD, TESTSUIES} The Execution Accuracy(EX) calculates the percentage of queries where the predicted SQL queries match the correct SQL queries when executed. Valid Efficiency Score(VES) measures the percentage of predicted SQL queries that output sets consisting of the results from the ground-truth SQL queries.

\subsection{Base Models}

In this paper, we adopt Qwen2.5-Coder-Instruct series as our base models, given their leading performance in the filed of code generation. We evaluate the effectiveness of our framework across multiple model sizes, including 1.5B, 3B and 14B, to demonstrate its ability to enhance lightweight models. To further explore the boundary of our method, we also conduct experiment using more powerful closed APIs, including GPT-4o-mini, GPT-4o and Gemini2.5. In future work, we aim to distill the reasoning precess into a lightweight model, which currently implemented through MCTS refinement.

\subsection{Hyper-parameters}
In order to ensure the stability of our experiment results, we standardized the hyper-parameters as follows. The temperature is fixed at 0.1, the top-p parameter is set to 1.0, and the max-token length is 32168. As for the hyper-parameters in the MCTS-Refiners, the child nodes of a node are set to 2, and the max-rollout numbers are 5.

\subsection{Experimental Results}

\subsubsection{A. BIRD Results} 
Table 1 presents a performance comparison of our method in the BIRD dataset against existing approaches. When using lightweight models(1.5B and 3B), MCTS-SQL outperforms ChatGPT-3.5 and even rivals some earlier methods based on GPT-4. This demonstrates that our method can be deployed on resource-constrained edge devices, without relying on large-scale models or costly APIs. Furthermore, when combined with the most powerful Gemini2.5-pro, MCTS-SQL reaching \textbf{72.91\%} execution accuracy (EX) and \textbf{72.66\%} value execution score (VES) on the development set, confirming its superiority over existing methods and its practical utility.

Table 2 shows the detailed performance across different complexity levels. Compared to the baseline, our method achieve significant improvements. Analyzing the results, we observe that the model improves more on simpler examples. This may because mistakes in these cases are mostly about syntax, and the MCTS-Refiner can fix them easily using its feedback-based editing process. However, with a stronger base model, MCTS can still bring clear improvements on harder examples by exploring different ways to fix the errors.

\begin{table}[htbp]
\centering
\begin{footnotesize} 
\begin{tabular}{lcccc}
\toprule
\textbf{Method} & \textbf{Simp.} &\textbf{Mod.} &\textbf{Chall.} &\textbf{All}\\
\midrule
Qwen-1.5B&15.36&14.96&9.78& 14.71\\
Qwen-3B&19.24&16.18&12.49& 17.68\\
\midrule
Ours + Qwen-1.5B&46.36&34.96&22.78&40.69\\
Ours + Qwen-3B&53.74&39.62&24.55&46.71\\
Ours + Qwen-7B&62.98&42.21&30.28&53.61\\
\midrule
Ours+GPT-4o-mini&68.56&57.76&45.83&63.15\\
Ours+GPT-4o&74.32&65.17&51.48&69.40\\
Ours+Gemini2.5-pro& 76.98 & 69.82&56.84 &72.91 \\
\bottomrule
\end{tabular}
\end{footnotesize}
\caption{Execution accuracy in BIRD development set. The Qwen models in this table are Qwen-Coder-Instruct.}
\end{table}

\subsubsection{B. Spider  Results}
Table 3 presents the performance comparison on the Spider dataset. When using lightweight models, our method achieves results that are already practically usable. Furthermore, when equipped with GPT-4o as the base model, MCTS-SQL achieves outstanding performance, reaching \textbf{89.17\%} on the development set and \textbf{88.74\%} on the test set. While existing approaches have already demonstrated strong results on this benchmark, our method continues to deliver highly competitive performance.

\begin{table}[htbp]
\centering
\begin{tabular}{lcc}
\toprule
\textbf{Method} &\textbf{EX(Dev)} &\textbf{EX(Test)} \\
\midrule
C3+ChatGPT & 81.80 &82.30\\
DIN-SQL+GPT-4 & 82.80 &85.30\\
DAIL-SQL+GPT-4 & 84.40 & 86.60\\
MAC-SQL+GPT-4 &86.75 &82.80\\
CHESS &87.2&-\\
MCS-SQL+GPT-4 &89.5 &89.6\\
XiYan-SQL & - & 89.65\\
\midrule
\textbf{Ours + Qwen-1.5B} &\textbf{67.45}&\textbf{71.68}\\
Ours + Qwen-3B& 74.03&73.98\\
\midrule
Ours+GPT-4o-mini&86.16 &83.74\\
Ours+GPT-4o&88.71&86.63\\
Ours+Gemeni2.5-pro &89.17& 88.74\\
\bottomrule

\end{tabular}
\caption{Execution accuracy on both dev and test set of spider. The Qwen models in this table are Qwen-Coder-Instruct.}
\end{table}

\begin{table}[htbp]
\centering
\begin{tabular}{cccc}
\toprule
\textbf{Setting} & \textbf{Avg Latency} &\textbf{Tokens} &\textbf{Ex Accuracy} \\
\midrule
single-shot & 0.63s & 197 &14.71\\ 
no-cache & 6.12s & 2274 & 40.69\\
with-cache & 2.84s & 864 &40.42\\
\bottomrule
\end{tabular}
\caption{Impact of Prefix-Cache on Latency, Token Usage, and Execution Accuracy}
\end{table}

\subsection{Prefix-Cache Effectiveness Evaluation}
To evaluate the effectiveness of the proposed prefix-cache mechanism, we conducted controlled experiments focusing on three key aspects: (I) inference latency, (II) token computation reduction, and (III) execution accuracy impact. All experiments were performed on the BIRD . For a fair comparison, we used the same base model (Qwen2.5-Coder-Instruct-1.5B) under two settings: with prefix-cache and without prefix-cache. The detailed results can be seen in Table 4.

The prefix-cache stores the intermediate key-value states for invariant parts of the input, such as database schema and few-shot examples. This enables the model to skip redundant computations during subsequent decoding steps. We measured inference time on 500 multi-turn SQL generation tasks.

This pre-computation, though highly efficient, means the model's sampling process begins from a fixed distribution. We think that may occasionally guide the model toward a sub-optimal SQL generation path, especially when integrating new, dynamic feedback from the suffix. However, we consider the minor performance is an acceptable trade-off.

\subsection{Ablation Study}

To evaluate the role of each component in MCTS-SQL, we perform an ablation study using Qwen-1.5B as the base model (Table 5). The results confirm that all modules are essential and reflect a core design principle: progressively reducing the search space to mitigate the limited reasoning capacity of lightweight models.

Removing the Selector lowers overall accuracy from 40.7\% to 36.2\%, as the model must handle the full schema with redundant tables and columns. This highlights the Selector’s importance in filtering irrelevant schema elements and simplifying prompts.Eliminating the Direct Generator further reduces accuracy to 34.8\%, since the refinement process lacks a strong starting point. By providing an initial candidate, the Direct Generator decreases the search depth required by the MCTS-Refiner.The MCTS-Refiner proves most critical: without it, accuracy drops sharply to 16.8\%, close to single-shot generation. This validates the necessity of iterative, feedback-driven refinement for correcting syntax and semantic errors.Finally, removing the Prefix-Cache has minimal impact on accuracy (40.7\% → 40.4\%) but significantly decrease the latency, demonstrating its role in improving efficiency without sacrificing performance.Overall, these findings show that each component contributes to an efficient and accurate Text-to-SQL process, enabling a lightweight model like Qwen-1.5B to deliver competitive performance under resource constraints.

\begin{table}
\centering
\begin{footnotesize}
\begin{tabular}{lcccc}
\toprule
\textbf{Setting} & \textbf{Simp.} &\textbf{Mod.} &\textbf{Chall.} &\textbf{All}\\
\midrule
Full Pipeline & 46.5& 35.0 & 22.8 &\textbf{40.7}\\
w/o Selector & 42.1 & 30.1 & 18.1 & 36.2\\
w/o Direct Generator & 39.9 & 29.8 & 17.5 & 34.8\\
w/o MCTS-Refiner &  19.8 & 13.5 & 8.2 & 16.8\\
w/o Prefix-Cache & 46.4 & 35.0 & 22.8 & 40.4\\
Single-shot  & 17.2 & 12.1 & 7.1 & 14.7\\

\bottomrule
\end{tabular}
\caption{Ablation study on BIRD dev set using Qwen-1.5B as base model.}
\end{footnotesize}
\end{table}



\section{Error Analysis}
We conducted an error analysis of the single-prediction model and found that 42\% of the errors were caused by syntax mistakes, wrong field selection, or misunderstanding of the schema. MCTS-SQL improves Text-to-SQL performance by using a trial-and-error feedback mechanism to guide the search for better SQL queries. Instead of relying on a single-shot generation process, it explores multiple candidate SQL queries and keeps those that are actually effective based on execution results. This allows the method to recover from some generation errors. The rate of syntax errors and errors in table or database selection has significantly decreased to 13\%, demonstrating the importance of our approach in addressing common pattern problems. We observe a change in error types during refinement: while syntax and simple field errors decrease rapidly, remaining failures are dominated by deep semantic or schema-linking issues. Specifically, MCTS is effective at correcting errors such as spelling mistakes, operator errors . However, it still struggles with errors caused by ambiguous natural language queries as well as complex multi-table joins involving foreign keys.

\section{Conclusion}
In conclusion, we propose \textbf{MCTS-SQL}, a novel framework to enhance Text-to-SQL performance of lightweight models. Our approach employs Monte Carlo Tree Search for iterative SQL refinement with three modules: Selector, Direct Generator, and MCTS-Refiner. To reduce the overhead of multi-step refinement, we design a token-level \textbf{prefix-cache mechanism}, which significantly improves inference efficiency. Experiments on SPIDER and BIRD show that, even with a 1.5B model, MCTS-SQL outperforms ChatGPT-3.5, and achieves competitive results with Gemini 2.5. These findings demonstrate that MCTS-SQL, together with prefix-cache, provides a practical and efficient solution for real-world applications under resource constraints. Overall, MCTS offers an approach that empowers lightweight models with stronger capabilities, providing an excellent baseline for other complex cross-domain research.

\bibliography{aaai2026}

\newpage
\appendix
\onecolumn

The appendix  presents some detailed information and experimental conclusions about the proposed MCTS-SQL.

\section{Details on experiments}

\subsection{An example of schema}

This appendix provides an example of the proposed semi-structured schema representation used throughout our framework. It demonstrates how table-level and column-level information is organized to translate the database structure clearly to the language model. For each table, we include its name and an optional description, followed by a list of column entries, each consisting of the column name, data type, description, and example values. Furthermore, explicit foreign key mappings are listed to illustrate the relationships between tables, enabling the model to reason over the database schema more effectively.

\begin{figure*}[htbp]
  \begin{center}
  \includegraphics[width= 3.5in]{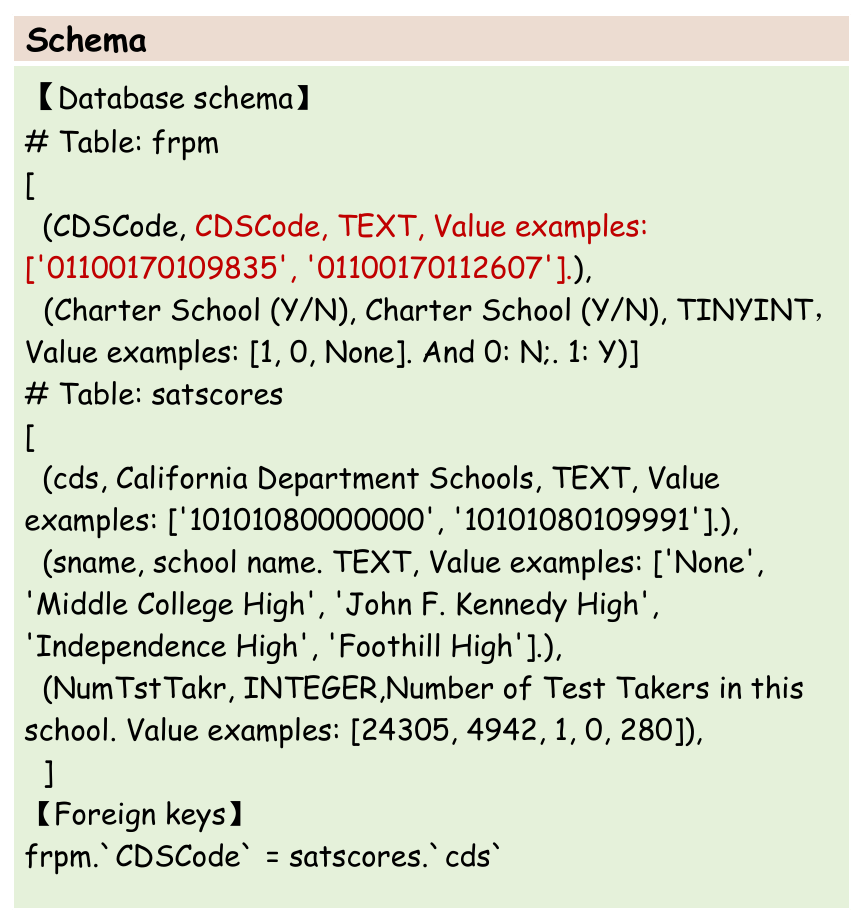}
  \caption{An example of proposed database schema format.The format consists of table names, descriptions and column level details(name, data type, description, and examples) to represent the hierarchical information of databases.}
  \label{Fig-appendix1}
  \end{center}
\end{figure*}

\subsection{Ablation study of hyper-parameters in MCTS}

This appendix section presents additional analysis on the sensitivity of two key hyper-parameters in our Monte Carlo Tree Search (MCTS) module: the maximum number of rollouts and the number of child nodes explored at each step. As shown in Table 1, we systematically vary these parameters and evaluate their effects on performance across different query difficulty levels. The results indicate that the overall performance is relatively stable under such changes, with only slight fluctuations observed. This demonstrates the robustness of our method to these hyper-parameter settings.

\begin{table}[htbp]
\centering
\begin{footnotesize}
\begin{tabular}{lcccc}
\toprule
\textbf{Pipeline} & \textbf{Simp.} & \textbf{Mod.} & \textbf{Chall.} & \textbf{All} \\
\midrule
\textbf{Ours + Qwen2.5-1.5B-coder} & \textbf{46.36} & \textbf{34.96} & \textbf{22.78} & \textbf{40.69} \\
\midrule
max-rollouts-5 & 46.36 & 34.96 & 22.78 & 40.69 \\ 
max-rollouts-6 & 46.12 & 35.02 & 22.51 & 40.53 \\
max-rollouts-7 & 46.44 & 34.61 & 22.93 & 40.64 \\
\midrule
child nodes-2 & 46.36 & 34.96 & 22.78 & 40.69 \\
child nodes-3 & 46.53 & 34.12 & 22.89 & 40.54 \\
\bottomrule
\end{tabular}
\caption{Ablation study of hyper-parameters in MCTS with EX on the development set using Qwen2.5-1.5B-coder.}
\end{footnotesize}
\end{table}

\subsection{An example of MCTS-Refine}

To demonstrate the effectiveness of our MCTS-Refiner, we provide an example. Starting from an initial incorrect SQL, the refiner iteratively explores and improves candidate queries by evaluating execution feedback and applying guided modifications. Through this process, it successfully constructs a correct SQL that joins relevant tables, applies appropriate filters, aggregates and sorts the results, and adds a constraint on the number of charter schools. This example highlights how MCTS-Refiner progressively resolves intent understanding, schema linking, and syntactic issues, ultimately producing an executable and accurate SQL query.

\begin{figure*}
  \begin{center}
  \includegraphics[width= 6in]{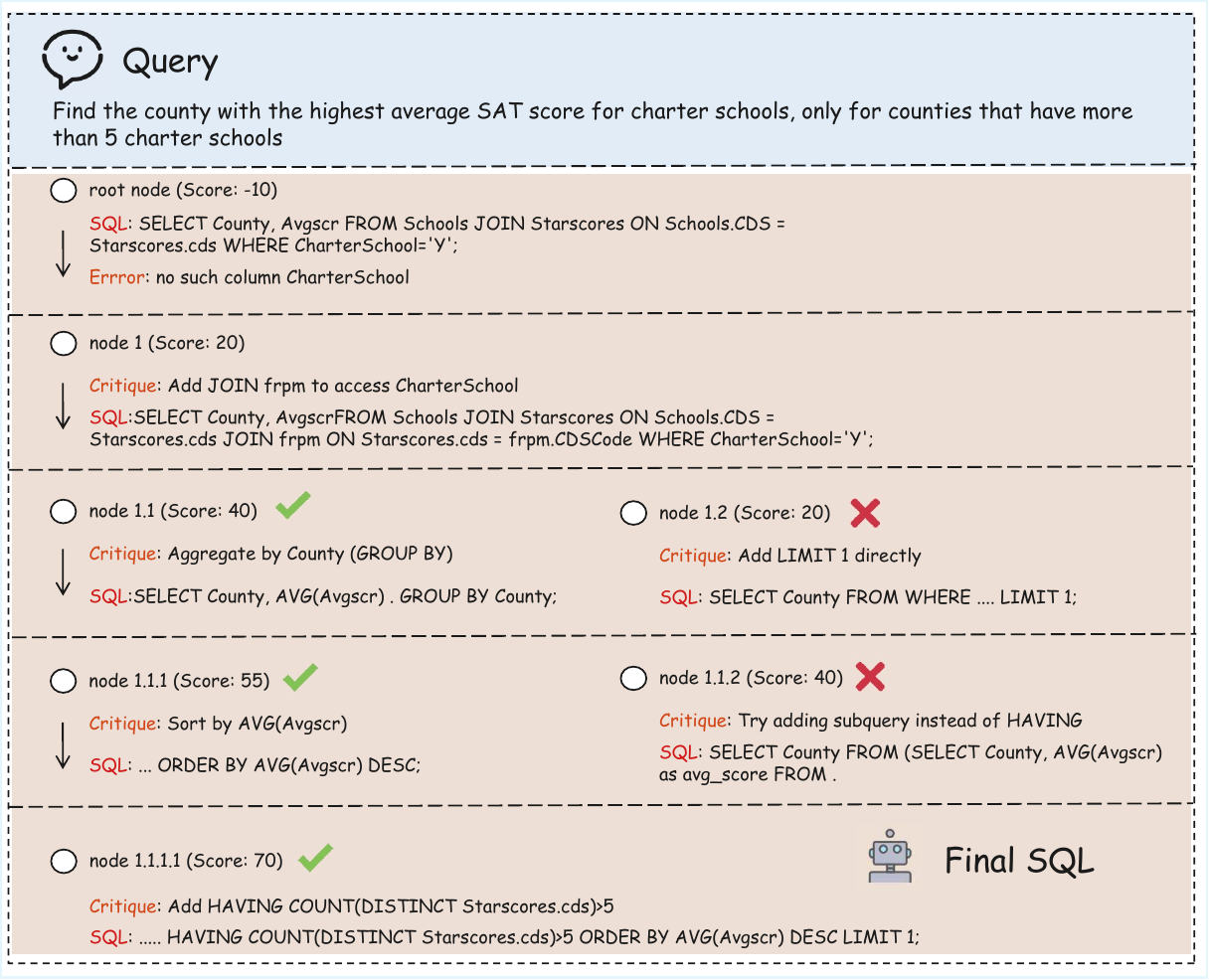}
  \caption{The figure illustrates how the MCTS-Refiner incrementally improves SQL generation by exploring candidate queries through a tree search. Starting from an initial faulty query, it applies guided edits based on execution feedback and critique until a correct SQL is formed that satisfies the input question.}
  \label{Fig-appendix2}
  \end{center}
\end{figure*}

\subsection{Baselines}

In this appendix, we provide a detailed description of the baselines used in the experiments presented in the paper. Baseline models serve as a benchmark for evaluating the effectiveness of our proposed approach, MCTS-SQL. The selection of these baselines is based on their relevance to the Text-to-SQL task and their ability to represent widely recognized methods in the field. Each baseline is briefly introduced, along with its specific configuration and the rationale for its inclusion in our study. The performance of these models is compared against that of our proposed framework to highlight the improvements achieved through MCTS-SQL.

\begin{itemize}

\item \textbf{DIN-SQL} utilizes zero-shot and few-shot prompting with large language models like GPT-4 and CodeX. While effective for basic queries, it often lags behind fine-tuned models on complex tasks and benchmarks like Spider due to its lack of fine-tuning.

\item \textbf{DAIL-SQL} enhances LLM-based Text-to-SQL by systematically comparing existing prompt strategies, optimizing question representation, and employing supervised fine-tuning. 

\item \textbf{MAC-SQL} introduces a multi-agent framework with three components: a Decomposer to break down complex queries, a Selector to focus on relevant database parts, and a Refiner to fine-tune the generated queries. This approach excels on more intricate queries by leveraging agent collaboration and fine-tuned models like SQL-Llama.

\item \textbf{MCS-SQL} employs multiple prompts to explore a wider search space, combined with schema linking and multiple-choice selection to filter and optimize generated SQL queries. This approach has set new benchmarks on the BIRD and Spider datasets.

\item \textbf{SQL-o1} leverages a self-reward-driven MCTS search mechanism to enhance reasoning in Text-to-SQL tasks. It includes multi-step exploration and dynamic pruning for faster inference, improving generalization and surpassing GPT-4 in performance on the BIRD dataset.

\item \textbf{Alpha-SQL} focuses on zero-shot Text-to-SQL using MCTS. It dynamically guides SQL construction actions with LLM-as-Action-Model, while using a self-supervised reward function to refine the SQL generation process.

\item \textbf{XiYan-SQL} uses an ensemble of multiple generators and fine-tuning strategies to improve SQL candidate diversity and quality. The approach integrates in-context learning (ICL) with M-Schema, a semi-structured schema representation, and includes a refiner to correct errors and select the best candidate.

\item \textbf{CHASE-SQL} adopts a multi-agent framework that generates diverse SQL candidates through a divide-and-conquer approach. It incorporates chain-of-thought reasoning, synthetic example generation, and pairwise candidate comparison, resulting in better SQL query generation and selection.

\item \textbf{DSAIR} focuses on metadata extraction techniques to support SQL generation, addressing challenges in database understanding. It explores profiling, query log analysis, and SQL-to-text generation, reducing the need for human-generated metadata and improving SQL generation quality.

\end{itemize}

While existing baselines have advanced the Text-to-SQL task through prompt engineering, agent collaboration, supervised fine-tuning, and search strategies, they often rely on either large-scale models or carefully designed training pipelines. In contrast, our motivation lies in enabling lightweight models to effectively handle complex SQL generation without large parameter counts or costly training. Unlike methods such as Alpha-SQL and SQL-o1 that use MCTS mainly in zero-shot or reward-driven settings, MCTS-SQL introduces a structured, feedback-guided refinement process tailored for small models, leveraging schema pruning, strong initialization, and efficient search. This design allows MCTS-SQL to strike a unique balance between performance and resource efficiency, distinguishing it from prior work that either emphasizes prompt diversity or model scale.

\section{Details of prompts}

\subsection{Prompt of selector}

\begin{tcolorbox}[colback=gray!10, colframe=black, boxrule=0.5pt, arc=3pt, fontupper=\ttfamily\small, breakable]
You are an expert of SQL. Given a detailed schema description, you should finish the task in the [instruction].

[DB\_ID]\{db\_id\}

[Schema]\{desc\_str\}

[Foreignkeys]\{fk\_str\}

Here are few examples:

\{schema describe\}

[Instruction] \\
Your responsibilities include:

1. Discard any table schema that is not related to the user question and accompanying evidence.

2. For each relevant table:

If the number of columns is 10 or fewer, mark the table as "keep-all".

Otherwise, select the top 6 most relevant columns and list them in descending order of relevance.

3. Ensure at least 3 tables are included in the final output.

4. Output must be in strict JSON format.

example's answers are: \\
\{example's answer\}

[Question] \\
\{query\}

[Evidence] \\
\{evidence\}

[Answer]

\end{tcolorbox}

\subsection{Prompt of direct generator}

\begin{tcolorbox}[colback=gray!10, colframe=black, boxrule=0.5pt, arc=3pt, fontupper=\ttfamily\small, breakable]
You are an expert of SQL. Given a detailed schema description, you should finish the task in the [instruction].

[DB\_ID]\{db\_id\}

[Schema]\{desc\_str\}

[Foreignkeys]\{fk\_str\}

Here are few examples:

\{schema describe\}

[Instruction] \\
Your responsibilities include:

1.Understand the schema and foreign key relationships to identify relevant tables and columns.

2.Carefully align the user question with the corresponding elements in the schema.

3.Resolve ambiguities based on schema structure and naming conventions.

4.Compose a complete and executable SQL query that correctly answers the user’s question.

5.Ensure the query respects SQL syntax and follows best practices (e.g., proper JOINs, WHERE conditions, GROUP BY, ORDER BY).

6. Do **not** output any explanation or commentary — only the SQL query.

7.If multiple interpretations exist, select the most plausible one based on provided evidence.

example's answers are: \\
\{example's answer\}

[Question] \\
\{query\}

[Evidence] \\
\{evidence\}

[Answer]

\end{tcolorbox}

\subsection{Prompt of verifier}

\begin{tcolorbox}[colback=gray!10, colframe=black, boxrule=0.5pt, arc=3pt, fontupper=\ttfamily\small, breakable]

You are an expert of SQL. Given a detailed schema description, you should finish the task in the [instruction].

[DB\_ID] \{db\_id\}

[Schema] \{desc\_str\}

[Foreignkeys] \{fk\_str\}

Here are few examples:

{schema describe}

[Instruction] \\
Your responsibilities include:
1.Understand the schema and table relationships based on the schema description and foreign key mappings.

2.Analyze the user question and determine what the expected result should be.

3.Review the given SQL query and check whether it correctly retrieves the expected result for the question.

4.Verify that the query uses the correct tables, columns, filters, aggregations, joins, groupings, and ordering.

5.Detect any mismatches, hallucinations (e.g., using non-existent columns), logic errors, or inconsistencies with the user intent.

6.If the SQL is correct, answer `YES`.  If the SQL is incorrect, answer `NO` and briefly explain the main error(s).

7.Do **not** rewrite or fix the SQL — only judge its correctness and explain why.

example's answers are: \\
\{example's answer\}

[Question] \\
\{query\}

[Evidence] \\
\{evidence\}

[SQL] \\
\{sql\}

[Answer]

\end{tcolorbox}

\subsection{Prompt of critique in MCTS}

\begin{tcolorbox}[colback=gray!10, colframe=black, boxrule=0.5pt, arc=3pt, fontupper=\ttfamily\small, breakable]

You are an expert of SQL. Given a detailed schema description, you should finish the task in the [instruction].

[DB\_ID] \{db\_id\}

[Schema] \{desc\_str\}

[Foreignkeys] \{fk\_str\}

[Instruction] \\
Your responsibilities include:
1. Identify specific parts of the SQL that are responsible for the error.

2. Check if any schema misunderstanding, incorrect column/table usage, invalid expressions, or logical flaws are present.

3. Provide **constructive and detailed** feedback to help improve the SQL, with reference to schema if necessary.

4. Do **not** rewrite or fix the SQL — only explain what's wrong and what to consider correcting.

[Question] \\
\{query\}

[Evidence] \\
\{evidence\}

[SQL] \\
\{sql\}

[SQL\_error]
{sql\_error}

[Answer]

\end{tcolorbox}

\subsection{Prompt of refine in MCTS}

\begin{tcolorbox}[colback=gray!10, colframe=black, boxrule=0.5pt, arc=3pt, fontupper=\ttfamily\small, breakable]

You are an expert of SQL. Given a detailed schema description, you should finish the task in the [instruction].

[DB\_ID] \{db\_id\}

[Schema] \{desc\_str\}

[Foreignkeys] \{fk\_str\}

[Instruction] \\
Your responsibilities include:

1.Continue to comply with the requirements in the original instructions.

2. Give a direct and concise SQL to the instructions, do not refer to or discuss the criticisms.

3. Do not repeat the original instruction or answer statements.

4.Response with only the answer.

[Question] \\
\{query\}

[Evidence] \\
\{evidence\}

[SQL] \\
\{sql\}

[SQL\_error]
{sql\_error}

[Critique]
{critique}

[Answer]

\end{tcolorbox}

\subsection{Prompt of evaluation in MCTS}

\begin{tcolorbox}[colback=gray!10, colframe=black, boxrule=0.5pt, arc=3pt, fontupper=\ttfamily\small, breakable]

You are an expert of SQL. Given a detailed schema description, you should finish the task in the [instruction].

[DB\_ID] \{db\_id\}

[Schema] \{desc\_str\}

[Foreignkeys] \{fk\_str\}

[Instruction] \\
Your responsibilities include:

1. Based on executing SQL and the error information, please provide a reward score between -100 and 100 for the anwer quality, using very strict standards.

2.Do not give a full score above 95. Make sure the reward score is an integer."

3.Return *ONLY* the score."

[Question] \\
\{query\}

[Evidence] \\
\{evidence\}

[SQL] \\
\{sql\}

[SQL\_error]
{sql\_error}

[Answer]
\end{tcolorbox}

\end{document}